\documentclass[12pt]{iopart}
\usepackage{epsfig}

\usepackage{iopams}
\usepackage{upgreek}
\usepackage{graphicx}
\usepackage[breaklinks=true,colorlinks=true,linkcolor=blue,urlcolor=blue,citecolor=blue]{hyperref}
\usepackage{cite}
\usepackage{longtable}
\usepackage{epstopdf}
\usepackage{xcolor}
\usepackage{colortbl,booktabs}

\begin{document}
	
	\title[Analysis of pulsed--DC SDBD actuators]{Modeling and theoretical analysis of SDBD plasma actuators driven by Fast-Rise-Slow-Decay Pulsed-DC voltages}
	
	\author[cor1]{Xiancong Chen$^1$, Yifei Zhu$^{1,2}$, Yun Wu$^{1,2}$, Zhi Su$^1$, Hua Liang$^1$}	
	\address{$^1$ Science and Technology of Plasma Dynamics Laboratory, Airforce Engineering University, Xi'an 710038, People's Republic of China}
	\address{$^2$ Institute of Aero-engine, School of Mechanical Engineering, Xi’an Jiaotong University, Xi’an 710049, People's Republic of China}
	\eads{\mailto{yifei.zhu.plasma@gmail.com, wuyun1223@126.com}}

	\begin{abstract}
	Surface dielectric barrier discharge (SDBD) actuators driven by the Pulsed-DC voltages are analyzed. The Pulsed-DC SDBD studied in this work is equivalent to a classical SDBD driven by a tailored Fast-Rise-Slow-Decay (FRSD) voltage waveform. The plasma channel formation and charge production process in the voltage rising stage are studied at different slopes using a classical 2D fluid model, the thrust generated in the voltage decaying stage is studied based on an analytical approach taking 2D model results as the input. 
	A thrust pulse is generated in the trailing edge of the voltage waveform and reaches maximum when the voltage decreases by approximately the value of cathode voltage fall~($\approx$600~V). The time duration of the rising and trailing edge, the decay rate and the amplitude of applied voltage are the main factors affecting the performance of the actuator. Analytical expressions are formulated for the value and time moment of peak thrust, the upper limit of thrust is also estimated. 
	Higher voltage rising rate leads to higher charge density in the voltage rising stage thus higher thrust. Shorter voltage trailing edge, in general, results in higher value and earlier appearance of the peak thrust. The detailed profile of the trailing edge also affects the performance. Results in this work allow us to flexibly design the FRSD waveforms for an SDBD actuator according to the requirements of active flow control in different application conditions.
	
	\end{abstract}

	\pacs{82.33, 52.20,  34.80}
	\vspace{2pc}
	\noindent{\it Keywords}: SDBD, pulsed--DC, thrust, fluid plasma model, analytical model
	
	\section{Introduction}
	Surface dielectric barrier discharges (SDBD) have been widely studied in aerodynamics community since 2000s~\cite{roth2000electrohydrodynamic,roth2003aerodynamic,enloe2004mechanisms,enloe2008time}. A typical SDBD device consists of a high-voltage electrode placed above the dielectric surface, and a low-voltage  electrode (typically grounded or at constant potential) placed below the dielectric surface. For a SDBD actuator driven by a sinusoidal voltage with voltage amplitude $\sim10\rm~kV$ and frequency $\sim10\rm~kHz$ at atmospheric pressure, micro-discharges appear stochastically in the vicinity of the high-voltage electrode and develop into streamers propagating along the dielectric surface, generating body force and inducing gas flow acceleration near the surface (also known as ionic wind). Some comprehensive reviews about plasma--fluid interaction can be found in \cite{Corkereview} and \cite{Leonov2016}, the mechanism of flow control by SDBD is thrust generation coupled with the external flow. When the high-voltage pulses are shorten to tens or hundreds of nanoseconds, a micro perturbation wave will be generated near the high-voltage electrode due to fast gas heating by the quenching of electronically excited states~\cite{Popov2011,Mintoussov2011,aleksandrov2010mechanism,takashima2012measurements,flitti_pancheshnyi_2009}. The features of ionic wind generation and fast gas heating are essential in  separation flow control~\cite{Roupassov2009,Thomas2009,Rethmel2011}, boundary-layer transition control~\cite{Grundmann2008,Humble2013} and skin-friction reduction in boundary/mixing layer~\cite{Timothy2006,Choi2010,Singh2019,Corke2018} et al.
	
	To deepen the understanding of SDBD in plasma--assisted flow control, and offer reference and basis for engineering design, some numerical investigations were also conducted in different conditions. Boeuf et al~\cite{boeuf2009contribution} presented a parametric study of the electro--hydrodynamic force generated by SDBD plasma actuators in air for sinusoidal voltage waveforms. Their 2D simulation results revealed that momentum was transferred from the charged particles to the neutral species in the same direction during both positive and negative parts of the cycle. The momentum transfer is due to positive ions during the positive part of the cycle, and to negative ions during the negative part of the cycle. Moreover, the contribution of negative ions tends to be dominant at low voltage frequencies and high voltage amplitudes. Soloviev~\cite{Soloviev_analyticalforceJPD,soloviev2015analytical} derived an analytical estimation for the thrust induced by a set of micro-discharges in SDBD configurations based on a phenomenological model. He confirmed Boeuf's perspectives~\cite{boeuf2009contribution} on the domination of negative ions in thrust generation, and further pointed out the origin of the force was the accumulation of volumetric negative charge carried by negative long--lived $\rm O_2^-$ and $\rm O_3^-$ ions. Above two studies all focused on the discharge properties in static air environments, recently Kinefuchi et al~\cite{Kinefuchi2018} numerically studied the shock--wave/boundary--layer separation control with two nSDBD configurations(parallel and canted with respect to the flow velocity vector) by large-eddy simulation (LES) and an energy deposition model for plasma actuator. A definite difference between the parallel and canted electrodes was founded: the former caused excess heating and increased the strength of the interaction, while the latter leaded to a reduction of the interaction strength, with a corresponding thinning of the boundary layer due to the momentum transfer.
	
	Recent experimental investigations reveal that, a SDBD plasma actuator driven by a pulsed-DC voltage waveforms can effectively reduce the skin friction drag reduction in turbulent boundary layers~\cite{McGowen2016,McGowan2016b,McGowan2018,Sontag2019,Thomas2019}. The so-called pulsed-DC SDBD plasma actuators have the same configurations with ordinary SDBD, but the applied voltage waveforms and its applying mode differ: an almost constant voltage was powered on the exposed electrode, while a voltage pulse with a sharp falling stage ($<20~\rm\mu s$) and a very slow rising stage ($\rm O(1)~\rm ms$) was powered on the buried electrode. These two different voltage waveforms were achieved by two DC sources, one for exposed electrode kept constant and the other for the buried electrode was periodically grounded for very short instants of time (DC--pulse width, $\rm O(20)~\rm\mu s$) by a solid--state switch. 
	The characteristics of the body--force--induced mean flow produced by the pulsed--DC actuator in static air were studied by Pitot probe and sample hot--wire measurements. The development of a transient velocity pulse showed approximately 6~m/s peak magnitude, and the time instant corresponding to the peak value coincided with the DC--pulse width. Moreover, the pulsed-DC actuator flow control arrays in the drag reduction experiments performed over the Mach number range of $0.05\leq M_{\infty}\leq0.15$ achieved unprecedented levels of drag reduction in excess of 70\%. Initial experiments of the acSDBD, nanosecond pulsed SDBD and pulsed-DC SDBD actuators have also been carried out in our group to study their effects on surface drag reduction on an airfoil, the results show that, a pulsed--DC SDBD device is more effective comparing with the other two actuators at the same voltage. Despite the impressive performance of pulsed--DC actuators in turbulence drag reduction, the underlying mechanisms are still unclear.
	
	The pulsed-DC SDBD can be considered as a classical SDBD device driven by a tailored voltage characterized by the Fast-Rise-Slow-Decay (FRSD) waveform on the exposed electrode. 
	The aim of this paper is to have a deeper understanding of the pulsed-DC SDBD (or the Fast-Rise-Slow-Decay SDBD, FRSDBD) actuator, in the view of discharge and thrust, by the combination of numerical and analytical approaches. A classical 2D fluid model is used to reveal the discharge properties at the voltage rising stage. An analytical model developed based on~\cite{Soloviev_analyticalforceJPD,soloviev2015analytical} is used to study the performance of the SDBD in the voltage decaying stage together with 2D results. 
		
	\section{Model description}
	The 2D PASSKEy (PArallel Streamer Solver with KinEtics) code is used. The code was used in modeling of nanosecond pulsed surface discharges~\cite{zhu2017nanosecond,zhu2018fgh,Zhu2019antiicing} and validated by measured discharge morphology, propagation velocity, voltage--current curves of experiments, and by a point-to-plane model benchmark~\cite{Kulikovsky1998}. Detailed mathematical formulations and validations can be found in paper~\cite{zhu2017nanosecond,Zhu2019antiicing}. In this section we briefly present the equations solved, and introduce the analytical model used for estimation of thrust in this work.
	
	\subsection{Fluid model formulation}
	The classical fluid model is used, drift-diffusion-reaction equations for species, Poisson equation for electric field, Helmholtz equations for photoionization are coupled. The drift-diffusion-reaction equations are:
	
	\begin{equation}\label{eqs.passkey.transport}
	\frac{\partial n_i}{\partial t} +\nabla\cdot\boldsymbol{\Gamma_i}=S_i+S_{ph}, i=1,2,...,N_{total}
	\end{equation}
	
	\begin{equation}\label{eqs.passkey.flux1} 
	\boldsymbol{\Gamma_i}=-D_i\nabla n_i - (q_i/|q_i|)\mu_in_i\nabla\Phi, i=1,2,...,N_{charge}
	\end{equation}
	\noindent
	where $\Phi$ is the electric potential, $n_{i}$, $q_{i}$, and $S_{i}$ is the number density, charge and source function for species $i$, respectively. The source function $S_{i}$ includes production and loss terms due to gas phase reactions and is calculated with detailed kinetics, and the kinetics scheme used in this paper has been validated in \cite{zhu2017nanosecond,zhu2018fgh}, $S_{ph}$ is the photoionization source term for electrons and oxygen ions. $D_{i}$ and $\mu_{i}$ are the diffusion coefficients and mobility of charged species, the electron swarm parameters and the rate coefficients of electron impact reactions are represented as explicit functions of the reduced electric field $E/N$ based on local field approximation (LFA). The diffusion coefficient and mobility for ions and other charged heavy species are founded from experimental data~\cite{Viehland1995}. In the code, $\nabla\cdot\boldsymbol{\Gamma_i}=0$ for neutral species is postulated. $N_{total}$ and $N_{charge}$ are the number of all species and charged species, respectively.
	
	Photoionization affects the propagation and morphology of the surface streamer. An efficient photoionization model based on three-exponential Helmholtz equations~ \cite{Bourdon2007,Luque2007} is used to calculate $S_{ph}$. We assume that the photoelectrons come from the ionization of oxygen molecules by VUV-radiation coming from electronically excited N$_2$ in $b^1\Pi_u$, $b$'$^1\Sigma^+_u$, $c$'$_4^1\Sigma^+_u$ states\cite{photo1982}.
	
	Poisson equation is solved for the entire computational domain:
	
	\begin{equation}\label{eqs.passkey.Poisson} 
	\nabla(\varepsilon_{0}\varepsilon\nabla\Phi)=-\sum_{i=1}^{N_{charge}}q_in_i-\rho_c
	\end{equation}
	\noindent
	where $\varepsilon_{0}$ is the permittivity of vacuum space and $\varepsilon$ the relative permittivity of the dielectric ($\varepsilon_{d}=4$) and air ($\varepsilon_{g}$=1.0), and $\rho_c$ is the surface charge density satisfying the continuting equation for charges on surfaces:
	
	\begin{equation}\label{eqs.passkey.chargecontin} 
	\frac{\partial \rho_c}{\partial t}=\sum_{j=1}^{N_{charge}}q_j[-\nabla\cdot\boldsymbol{\Gamma_j}]
	\end{equation}
	
	The propagation length of the streamer is an important parameter necessary for the following analytical model. When the voltage rising time is very short (no more than tens of nanoseconds) the calculated streamer could significantly exceed the reasonable value observed in the experiments, at such conditions an ``ion sink'' term has to be added to the continuity equation of ions\cite{Soloviev2014}. For cases with longer voltage rising time like in this case, the ions have enough time to dissipate and the correction term is no longer needed. We have conducted two test cases and confirmed this point.
		
	\subsection{Analytical model}
	The momentum source due to charged particle collisions with neutral gas molecules (thrust) can be expressed as~\cite{Boeuf2005,Soloviev_analyticalforceJPD}:
	
	\begin{equation}\label{eqs.bodyforce1}
	F=e(n_p-n_i-n_e)E_{xl}
	\end{equation}
	\noindent
	where $n_p$, $n_i$ and $n_e$ are the number density of positive ions, negative ions and electrons, respectively, $e$ is the elementary charge and $E_{xl}$ is the electric field in $x$ direction (the flow direction). In this paper, we only take the $x$--component of body force into account because the $y$--component of that is very small. The integration of body force over space and its averaging over time give the thrust per unit electrode length~\cite{Soloviev_analyticalforceJPD,soloviev2015analytical}:
	
	\begin{equation}\label{eqs.bodyforce2}
	F_{mean}=\frac{1}{T_v}\int_0^{T_v}\int_{\Omega}f(t,x,y)\mathrm{d}x\mathrm{d}y\mathrm{d}t
	\end{equation}
	\noindent
	where $T_v$ is the period of the applied voltage.
	
	Previous works reveal that the main contribution to momentum source occurs at when the slope of applied voltage is negative; body force is primarily due to the accumulation of volumetric negative charge, the main origin is $\rm O_2^-$~\cite{boeuf2009contribution,Soloviev_analyticalforceJPD}. After the end of a discharge, the electrons decay mainly by three--body electron attachment process:
	
	\begin{equation}\label{eqs.threebodyattach}
	\rm e+O_2+O_2\rightarrow O_2^-+O_2
	\end{equation}
	The residence time of negative ions inside the discharge zone can be characterized by:
	
	\begin{equation}\label{eqs.residence_time}
	\Delta\tau_q\approx l/V_{dri}=l/\mu_i E_{xl}
	\end{equation}
	where $\mu_i$ is the mobility of ions, in this work we use the same ion drift velocity $V_{dri}=100~\rm m\cdot s^{-1}$ as Ref~\cite{Soloviev_analyticalforceJPD,soloviev2015analytical}, the value is chosen based on the numerical simulation~\cite{Soloviev2009}. The accumulated negative charge per unit electrode length obeys the equation~\cite{Soloviev_analyticalforceJPD}:
	
	\begin{equation}\label{eqs.charge_evolution}
	\frac{\mathrm{d}q_n(t)}{\mathrm{d}t}\approx-\frac{q_n(t)}{\Delta\tau_q}
	\end{equation}
	The term on the right--hand side of above equation is the sink term due to negative ions drifting onto the dielectric surface. The solution of equation~(\ref{eqs.charge_evolution}) with initial condition $q_n(0)=q_0$ reads
	
	\begin{equation}\label{eqs.charge_evolution2}
	q_n(t)\approx q_0\mathrm{exp}\left(-\frac{t}{\Delta\tau_q}\right)
	\end{equation}
	\noindent
	where $q_0$ is the the total charge of electrons at the beginning of the thrust generation stage, obtained from the classical fluid modeling. The $x$--component of electric field $E_{xl}$ is represented as follows~\cite{Soloviev_analyticalforceJPD,soloviev2015analytical}:
	
	\begin{equation}\label{eqs.field}
	E_{xl}=\frac{\varepsilon V_s}{2l}
	\end{equation}
	where $V_s$ is the potential of dielectric surface charge layer, $l$ is the discharge length, $\varepsilon$ is the relative permittivity of the dielectric. Substituting formula~(\ref{eqs.field}) into equation~ (\ref{eqs.bodyforce1}), the thrust at time instant $t$ equals to
	
	\begin{equation}\label{eqs.bodyforce3}
	F=q_n(t)E_{xl}(t)=q_n(t)\frac{\varepsilon V_s(t)}{2l}
	\end{equation}
    Substituting equation(~\ref{eqs.bodyforce3}) into (\ref{eqs.bodyforce2}) we can obtain the time--averaged thrust.
	
	\subsection{Geometry, initial/boundary conditions}
	\label{sec:Geometry, initial/boundary conditions}
	
	\begin{figure}[t!h!]
		\epsfxsize=\columnwidth
		\begin{center}
			\epsfbox{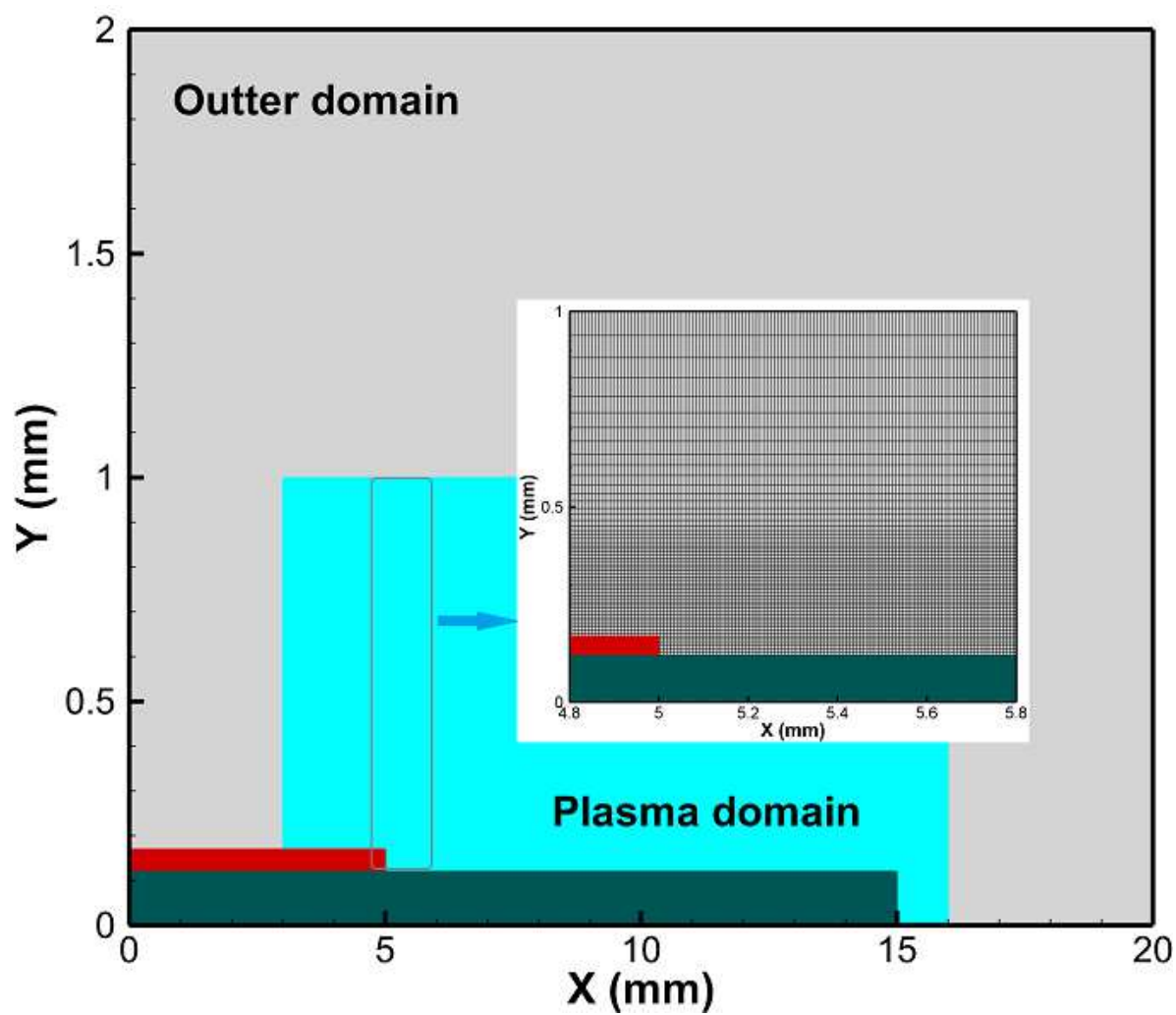}	
			\caption{Geometry, computational domain and mesh distribution (units in mm) for different equations. Exposed electrode: red domain; The dielectric: dark green domain. Transport equations: light blue domain (plasma domain); Poisson equation and Helmholtz equations: entire domain.}
			\label{fig:domain}
		\end{center}
	\end{figure}
    
    The studied SDBD actuator in the experiment conducted by our group has a typical configuration as mentioned in the introduction: a low-voltage electrode located at $y=0$, the dielectric (Kapton film, the relative permittivity $\varepsilon$ is 4) with $120~\rm\mu m$ thickness and a exposed electrode with $50~\rm\mu m$ thickness. The discharge experiment was operated in atmospheric air at ambient temperature. The geometry, air pressure and temperature used for PASSKEy code are same as these in the experiment. A computational domain of size $\rm 3~cm\times3~cm$ is assigned, the reduced computational domain and refined mesh distribution is shown in figure~\ref{fig:domain}. An uniform mesh size of $5~\rm\mu m$ is assigned for plasma domain, beyond the plasma domain the mesh size grows exponentially until the end of the entire computational domain.
    
   	\begin{figure}[t!h!]
   	\epsfxsize=\columnwidth
   	\begin{center}
   		\epsfbox{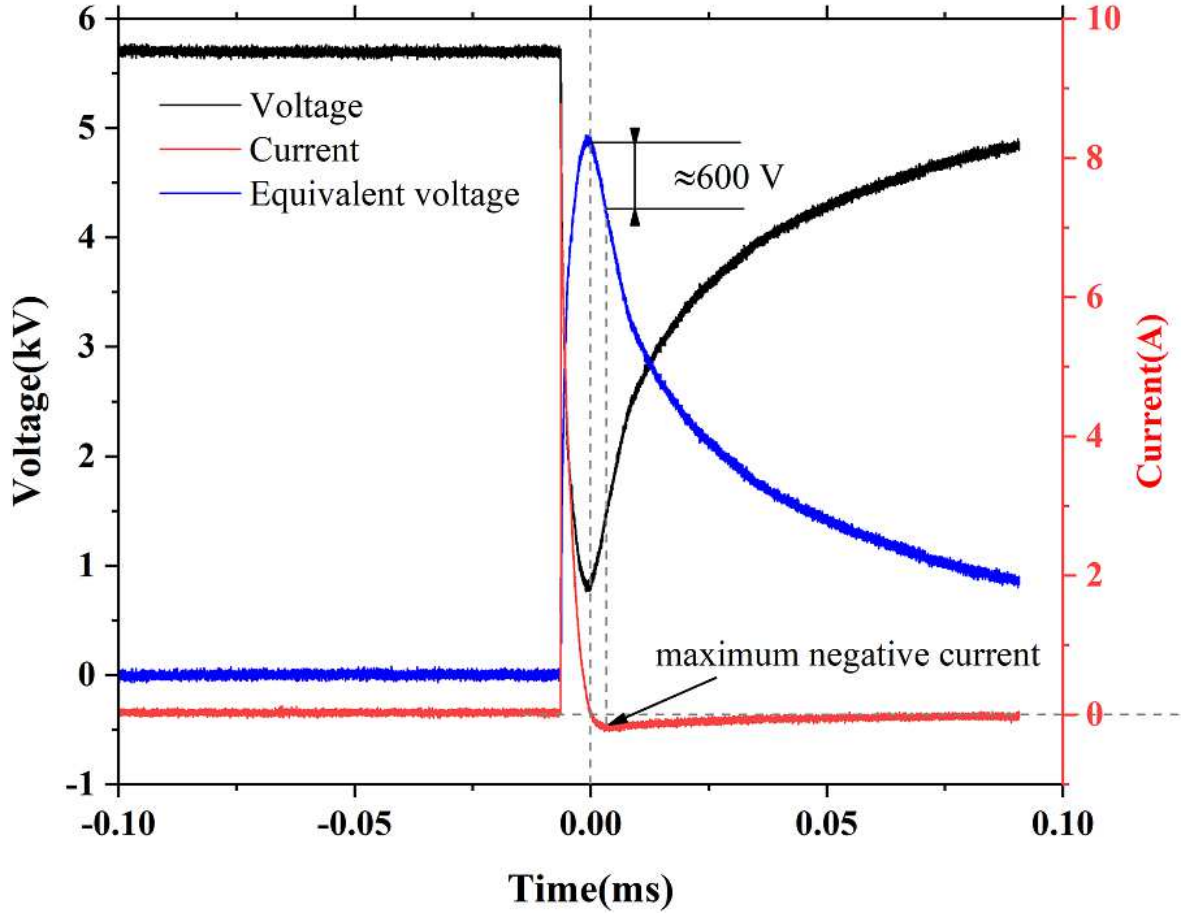}	
   		\caption{The experimentally measured voltage (for low--voltage electrode) and current waveforms. The ``Equivalent voltage'' refers to the potential gap between high--voltage and low--voltage electrodes.}
   		\label{fig:voltage}
   	\end{center}
   \end{figure}

    The initial condition is given by setting a background electron density of $n_{e0}=10^4~\rm cm^{-3}$ across the entire plasma domain. The initial ions density is set based on quasi-neutrality. The details of the boundary conditions of the Poisson equation, Helmholtz equations and transport equations in the PASSKEy code are described in~\cite{zhu2017nanosecond,zhu2018fgh}. The boundary conditions of Poisson equation for two metallic electrodes are Drichlet boundary conditions given by powered voltages on them. The experimental voltages are used as the input to PASSKEy code. The measured voltage applied on low-voltage electrode is shown in figure~\ref{fig:voltage} (the black line), and the high-voltage electrode is powered by constant voltage 5.7~kV.
    
    Our simulations have confirmed that, the discharge dynamics, the discharge currents, the breakdown voltages and their corresponding time instants are totally the same in the following two configurations of SDBD:
    
    Configuration (i): both electrodes are powered by DC voltages with the buried electrode periodically grounded to generate a pulse. The term ``pulsed-DC'' comes from this experimental configuration.
    
    Configuration (ii): only the exposed electrode is powered by the ``Equivalent voltage'' (the so called tailored FRSD voltage) drawn in figure~\ref{fig:voltage} (blue line) while the buried electrode is grounded as in classical SDBD configurations. 
    
    The equivalence of above configurations simplifies our modeling and theoretical analysis. Discussions in the following sections are all based on the configuration (ii), i.e. the ``voltage rising stage'' in configuration (ii) is equivalent to the ``voltage decaying stage'' in configuration (i).
        
    \section{Results and discussion}
    
    \subsection{Formation of plasma channel in voltage rising stage}
	\label{sec:Discharge dynamics}

   	\begin{figure}[t!h!]
	\epsfxsize=\columnwidth
	\begin{center}
		\epsfbox{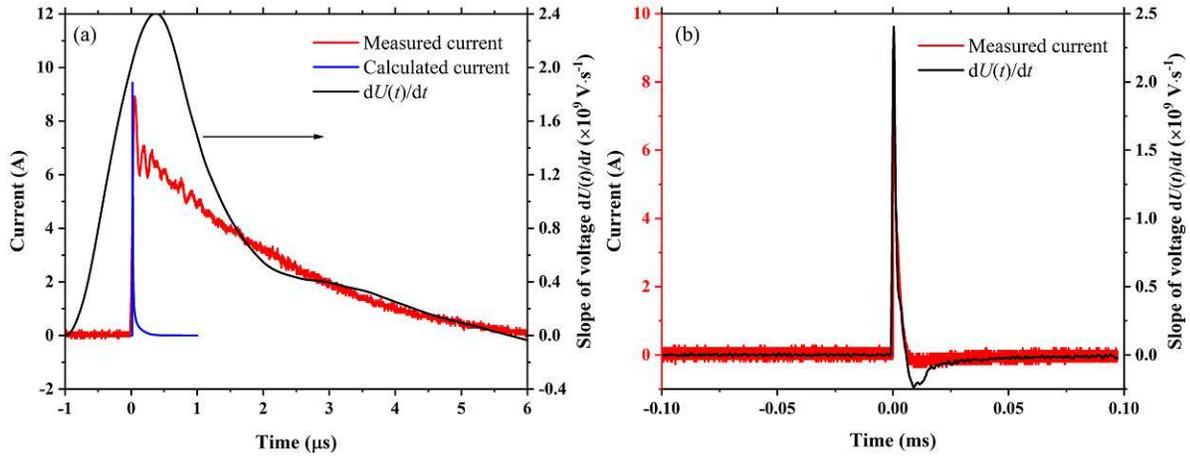}	
		\caption{Temporal profiles of the measured and calculated current, and the derivative of measured applied voltage $\mathrm{d}U(t)/\mathrm{d}t$ in time scale of (a) 7~$\mu$s and (b) 0.2~ms. The derivative is determined by smoothed measured voltage.}
		\label{fig:current}
	\end{center}
	\end{figure}
        
    Temporal evolution of measured and calculated discharge current is presented in figure~\ref{fig:current}(a). The breakdown takes place at applied voltage of 1040~V with a sharp increase of current value, the peak calculated current is 9.43~A, close to the measured value 8.76~A. The primary difference between the calculated and measured currents are the temporal profiles after the peak moment. The slowly decaying part in experimental current is the displacement current (which is not calculated in the code), the formula to characterize displacement current density can be written as follows:
    
    \begin{equation}\label{eqs.displacement}
    \boldsymbol{J}=\boldsymbol{j}_c+\varepsilon_{0}\frac{\partial\boldsymbol{E}}{\partial t}
    \end{equation}
    \noindent
    where $\boldsymbol{J}$ is total current density, $\boldsymbol{j}_c$ is the conductive current density and $\varepsilon_{0}\partial\boldsymbol{E}/\partial t$ is the displacement current. Here we calculate $\mathrm{d}U(t)/\mathrm{d}t$ as a rough estimation of $\mathrm{d}\textbf{E}/\mathrm{d}t$ assuming that $|\textbf{E}|$ is approximately proportional to the applied voltage $U(t)$ ($|\textbf{E}|\propto U(t)$). In other words, the value of the displacement current is estimated by the derivative of applied voltage $\mathrm{d}U(t)/\mathrm{d}t$. 
    We plot the estimated ``displacement current'' in figure~\ref{fig:current}(a) and (b) and compare the profile with the measured current in the decay region. The overlapping of the profiles of $\mathrm{d}U(t)/\mathrm{d}t$ and current values confirms our guess. Actually, in Ref~\cite{Thomas2019}, a similar experiment on a pulsed--DC SDBD actuator with the same voltage frequency and similar waveform and trailing edge was conducted, a current spike was found to occur on a much shorter time scale $\rm O(10^{-7})~\rm s$ comparing with DC--pulse width ($20~\rm\mu s$). Our calculated current spike occurs in time range 490~ns, qualitatively agrees that measured in Ref~\cite{Thomas2019}.
    	
    \begin{figure}[t!h!]
    	\epsfxsize=\columnwidth
    	\begin{center}
    		\epsfbox{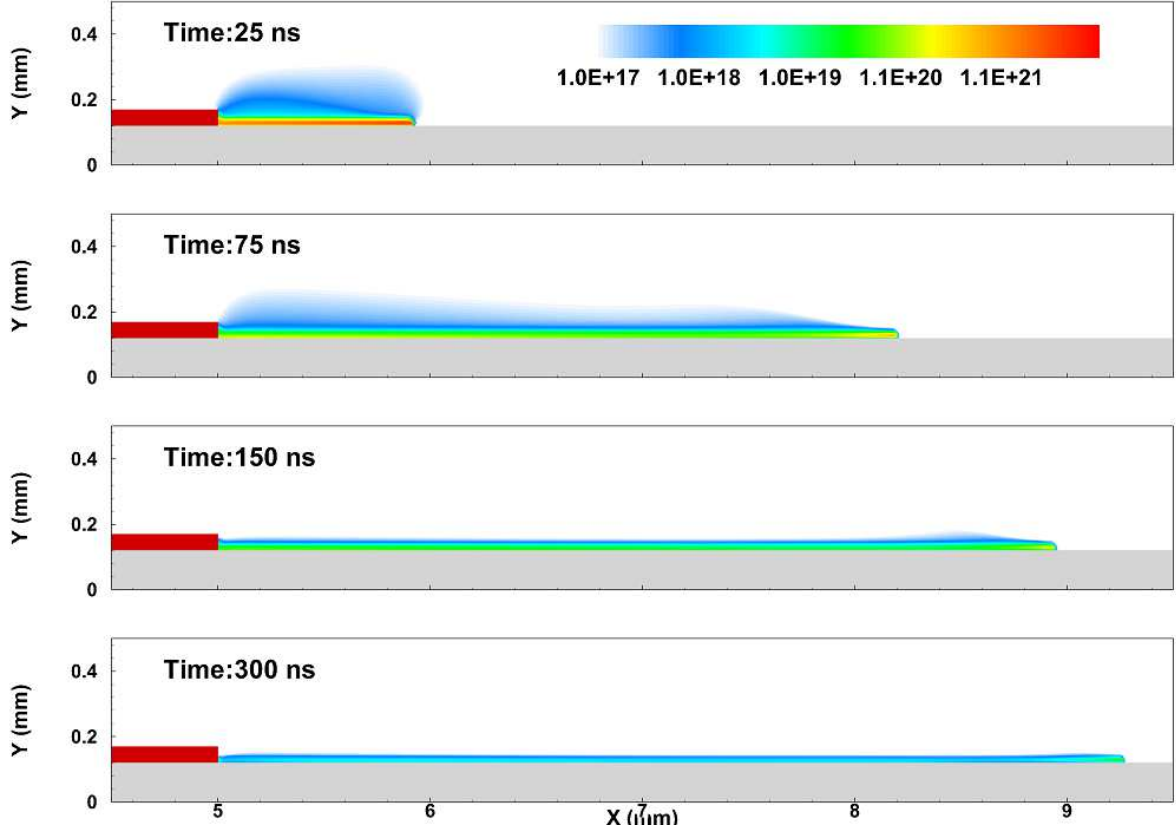}	
    		\caption{The evolution of electron density at time instant 25, 75, 150 and 300~ns (units in $\rm m^{-3}$).}
    		\label{fig:ne2d}
    	\end{center}
    \end{figure}

	The spatial evolution of electron density at time instants 25, 75, 150 and 300~ns are shown in figure~\ref{fig:ne2d}. The streamer has propagated 0.9~mm at 25~ns and a plasma channel with the thickness of about 50~$\rm\mu m$ is formed. The electron density in the channel is on the magnitude of $10^{18}~\rm m^{-3}$, with a higher level of about $10^{21}~\rm m^{-3}$ in the region about 16~$\rm\mu m$ above the dielectric surface. The total propagation distance is 4.3~mm, close to the analytical estimation 4.5~mm according to~\cite{Soloviev_analyticalSDBD}. The electron density in the channel decays when $t>25~\rm ns$, results in the decrease of conductive current, as shown in figure~\ref{fig:current}(a) and it decreases to a level below $10^{18}~\rm m^{-3}$ within 275~ns. 
	
	\begin{figure}[t!h!]
		\epsfxsize=\columnwidth
		\begin{center}
			\epsfbox{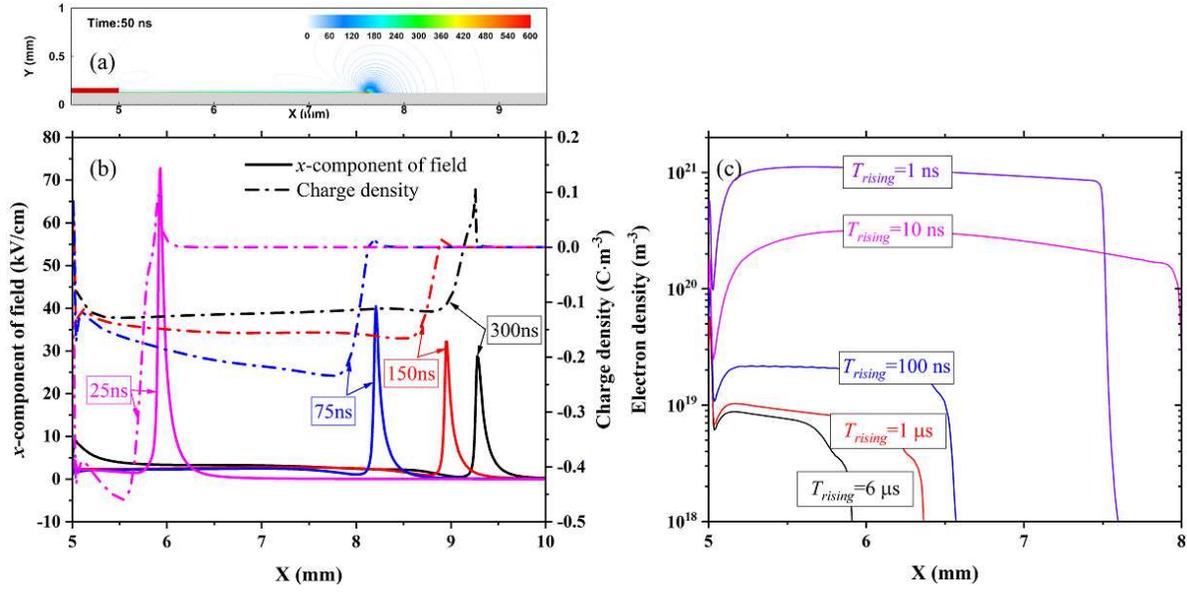}	
			\caption{(a)~The reduced electric field (units in Td) at time instant 50~ns; (b) Spatial profiles of $x$--component of electric field $E_{xl}$ and charge density located at horizontal line $y$=150~$\rm\mu m$ (30~$\rm\mu m$ above the dielectric surface) at time instants 25, 75, 150 and 300~ns. (c) Spatial profiles of electron density located at horizontal line $y$=150~$\rm\mu m$ for 5 cases with rising edges 6.8 and 1~$\rm\mu s$, 100, 10 and 1~ns, these data corresponding to time moments when the maximum electron density is reached in the channels.}
			\label{fig:enxcharge}
		\end{center}
	\end{figure}

    To study the thrust during the voltage rising stage, 2D-plot of electric field at 50~ns, as well as the spatial distribution of $x$--component of electric field $E_{xl}$ and charge density located at horizontal line $y$=150~$\rm\mu m$ at four successive time instants are presented in figure~\ref{fig:enxcharge}(a) and (b). The charge density along the line is negative except in the region at the front of the streamer head, while $E_{xl}$ in the whole channel is positive. The opposite signs between $E_{xl}$ and charge density means the thrust is negative and rather short in time scale. In other words, the thrust in the voltage rising stage does not contribute to flow control.
    
    The influence of voltage rising time on the discharge properties and thrust are also of interest. Here we conducted a parametric calculation of different voltage rising time at a fixed voltage amplitude (5~kV used in this work): 6.8~$\rm\mu s$ (experimental voltage in this work), 1~$\rm\mu s$, 100, 10 and 1~ns. The results are shown figure~\ref{fig:enxcharge}(c). The decreased rising edges shorten the time for the voltage to reach the breakdown field and increase the electric field in streamer heads in the same moment. During the streamer development stage, the electrons are absorb into the ionization head while the positive ions remain behind, creating a dipole with the characteristic length $1/\alpha_T$ ($\alpha_T$ is Townsend ionization coefficient) and electron density $n_e(x)=n_{e0}\mathrm{exp}(\alpha_T x)$ ($x$ is the distance from high-voltage electrode), the value of $n_e$ in on the order of $10^{18}\sim10^{21}~\rm m^{-3}$ ~\cite{fridman2008plasma}. The elevated streamer head field at a shorter rising slope results in higher Townsend ionization coefficient, thus higher electron density: at experimental rising edge the electron density is on the order of $10^{19}~\rm m^{-3}$, when the rising time is reduced to 1~ns the produced electron density grows by two orders of magnitude, to $10^{21}~\rm m^{-3}$. 
    
    According to equation~(\ref{eqs.charge_evolution2}) and (\ref{eqs.bodyforce3}), the thrust is proportional to initial charge, therefore higher thrust could be achieved in case of higher initial charge density using shorter voltage rising time. A significantly high charge density may shield the electric field inside the the plasma channel, but the thrust is generated mainly in the voltage decaying stage in time scales of microseconds, at this time scale the electrons have already attached to the molecules and atoms and the shielding effects are strongly weaken. 
    
    We note that, due to the short pulse, there is a gas heating process leading to a weak shock wave. The gas heating effects of SDBD is more significant in nanosecond time scales, the mechanisms and efficiency of gas heating energy production and how the heated gas interact with the ambient gas have been discussed in plenty of previous publications~\cite{Starikovskii2009,Popov2011,aleksandrov2010mechanism,Zhu2013Modelling,zhu2018fgh}, thus in this work, we do not focus on the gas heating and corresponding wide--known micro shock wave phenomena. 
    
    \subsection{Generation of the thrust in the voltage decaying stage}
    \label{Generation of thrust in the DC stage}

    We have known that the charge density in the plasma channel is negative, thus the positive thrust can only be generated when the $x$--component electric field $E_{xl}$ changes its direction. As shown in figure~\ref{fig:voltage}, the current changes its sign when the applied voltage reaches its peak value about 5000~V and the current decreases to 0, this reversion indicates $E_{xl}$ has changed its sign, positive thrust starts. 
    
    The negative current keeps increasing until it reaches its maximum value at 0.03~ms when the the potential gap between the high-voltage electrode (applied voltage $V$) and dielectric surface charge layer (potential $V_s$) reaches maximum $|V-V_s|\approx600~\rm V$, note that 600~V coincides with the value of cathode fall $\Delta V_c$ defined in Ref~\cite{Soloviev_analyticalforceJPD}. We denote the time moment of peak negative current as $t_{peak}$, then if $t>t_{peak}$, the current value will drop as the charge keeps following equation~(\ref{eqs.charge_evolution}).
        
    The potential of dielectric surface charge layer, $V_s$, is the key value to calculate the thrust of SDBD according to equation~(\ref{eqs.bodyforce3}). If $t>t_{peak}$, we will have $V_s=V-\Delta V_c\approx V-600$. The applied voltage $V$ can be expressed as the sum of two exponential functions:
    
    \begin{equation}\label{eqs.voltage}
    V(t)=\alpha V_m\mathrm{e}^{-\frac{\mathrm{ln}(\alpha V_m)t}{\beta T_{trailing}}}+(1-\alpha)V_m\mathrm{e}^{-\frac{\mathrm{ln}((1-\alpha)V_m)t}{T_{trailing}}}
    \end{equation}
    \noindent  
    where $V_m$ is the maximum value of applied voltage, $\alpha$ and $\beta$ ($\alpha, \beta<1$) are two parameters representing how fast the voltage drops, $T_{trailing}$ is the time duration of the trailing edge of applied voltage determing by the sub-voltage falling more slowly (the latter term in equation~(\ref{eqs.voltage})). These 4 parameters in equation~(\ref{eqs.voltage}) can be determined by fitting the measured voltage waveform. We summarize in table~\ref{tab:parameters} the values of these parameters for the voltage profiles used in this work (the one shown in figure~\ref{fig:voltage}) and in \cite{Thomas2019}. If $t<t_{peak}$, the thrust in this short time scale can be simplified as a linear function. Combining equation~(\ref{eqs.charge_evolution2}), (\ref{eqs.bodyforce3}) and (\ref{eqs.voltage}), the thrust $F(t)$ (units in $\rm{N\cdot m^{-1}}$) in the voltage decaying stage can be written as a piecewise function of time:
        
    \begin{equation}\label{eqs.thrust}
    F(t)=\left\{
    \begin{array}{lll}
    F_{peak}t/t_{peak} & & {t \leq t_{peak}}\\
    \varepsilon q_0(V(t)+600)\mathrm{exp}(-t/\Delta\tau_q)/2l & & {t > t_{peak}}
    \end{array} \right.
    \end{equation}
    
    \begin{equation}\label{eqs.thrust_max}
    F_{peak}=\varepsilon q_0V_m\mathrm{exp}(-t_{peak}/\Delta\tau_q)/2l
    \end{equation}
    
    \begin{table}[htbp]  \caption{Parameters of formula~(\ref{eqs.voltage}) describing the applied voltages.}
    	
    	\label{tab:parameters}
    	\centering
    	\begin{tabular}{ccccl}  
    		\toprule   
    		
    		$V_m$~(kV) & $T_{trailing}$~(ms) &$\alpha$ & $\beta$ & Reference \\  
    		
    		\midrule   
    		
    		5 & 0.6 & 0.4662 &0.1345& This work  \\  
    		
    		4 &  1.92 &0.4685 &0.2623 & \cite{Thomas2019} \\    
    		
    		\bottomrule  
    		
    	\end{tabular}
    	
    \end{table}
    
    Assuming all charges of electrons was converted to negative ions by three-body attachment process~(\ref{eqs.threebodyattach}). The initial charge density can be estimated as $q_0\approx e\cdot n_{emax}\cdot l_t\cdot\delta=1.6\cdot10^{-19}\times10^{21}\times0.9\cdot10^{-3}\times16\cdot10^{-6}=2.3\cdot10^{-6}~\rm C\cdot m^{-1}$ ($e$, $n_{emax}$, $l_t$ and $\delta$ represent the elementary charge, electron density, discharge length, and the thickness of region with magnitudes of electron density $n_{emax}$ at 25~ns, respectively, values were obtained by 2D calculation). $l=4.3~\rm mm$ is the total discharge length obtained in 2D model, $\Delta\tau_q=l/V_{dri}=4.3\cdot10^{-5}~\rm s$.
    
    \begin{figure}[t!h!]
    	\epsfxsize=\columnwidth
    	\begin{center}
    		\epsfbox{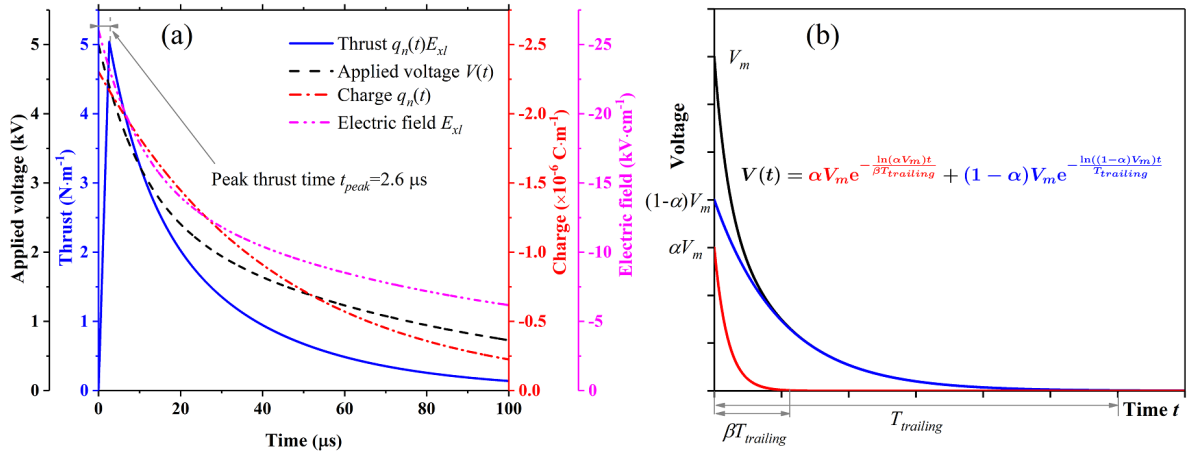}	
    		\caption{(a) Temporal profiles of thrust, applied voltage, charge and $x$--component of electric field. These profiles corresponding to trailing edge $T_{trailing}=0.6~\rm ms$, $V_m=5~\rm kV$, $q_0=2.3\cdot10^{-6}~\rm C\cdot m^{-1}$. (b) The schematic diagram of voltage characterized by equation~(\ref{eqs.voltage}), $T_{trailing}$ and $\beta T_{trailing}$ is determined by the time moment when $V(t)=1$~V.}
    		\label{fig:param4}
    	\end{center}
    \end{figure}

    Temporal profiles of thrust $q_n(t)E_{xl}$, applied voltage $V(t)$, charge $q_n(t)$, $x$--component of electric field for our experimental voltage are plotted in figure~\ref{fig:param4}. The thrust reached its peak value at 2.6~$\rm\mu s$ and then decreased to below 1~$\rm N\cdot m^{-1}$ within 40~$\rm \mu s$. The evolution of thrust after the peak time was dominated by $q_n(t)$ due to the fast charge decay rate. For example, when $t=100~\rm\mu s$, $q_n(t)$ had decreased by $1-e^{-100/43}\approx 90\%$ of the initial charge according to equation~(\ref{eqs.charge_evolution2}).     
    
    \begin{figure}[t!h!]
	\epsfxsize=\columnwidth
	\begin{center}
		\epsfbox{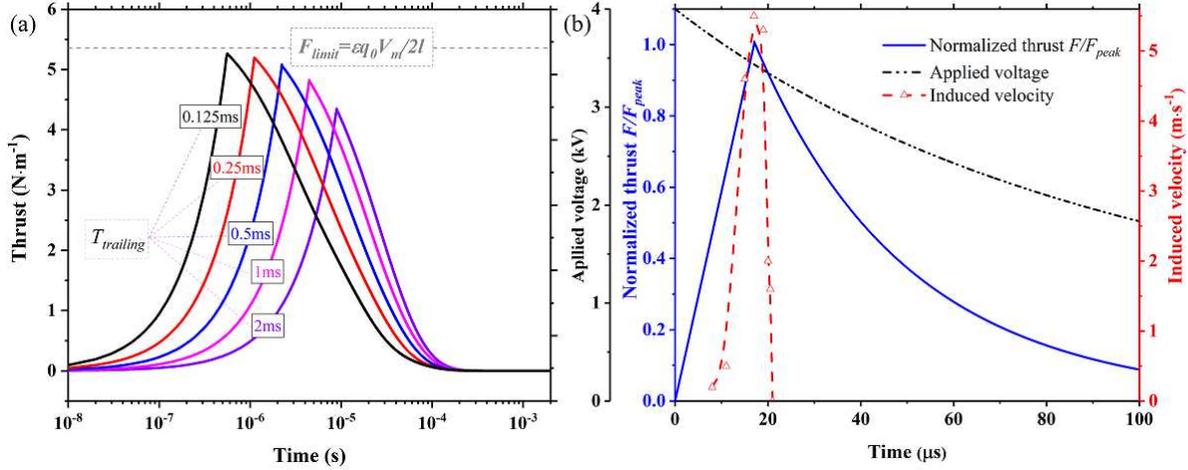}	
		\caption{(a) Temporal profiles of thrust when $T_{trailing}=$ 2, 1, 0.5, 0.25, and 0.125~ms. $F_{limit}$ is the theoretical limit of thrust under the identical condition. (b) Temporal profiles of calculated normalized thrust $F(t)/F_{peak}$ derived from equation~(\ref{eqs.thrust}). Applied voltage and the sample hot--wire measurements of the transient velocity near the surface produced by the pulsed--DC actuator from~\cite{Thomas2019} are plotted as a reference. Parameters in equation~(\ref{eqs.voltage}) describing the voltage waveforms in~\cite{Thomas2019} are shown in table~\ref{tab:parameters}.}
		\label{fig:force}
	\end{center}
	\end{figure}
    
    Based on the 2D calculation and equation~(\ref{eqs.thrust}), the most interesting questions concerning the pulsed--DC SDBD (or the Fast-Rise-Slow-Decay SDBD, FRSDBD in this work) can be answered:
    
    (1) The mechanism of thrust generation. The thrust is generated due to the motion of negative charged ions in the voltage decaying stage. Different from traditional SDBD actuator driven by sinusoidal voltage waveforms, the charge is produced in only one pulse. The transient thrust $F(t)$ generated at different voltage trailing edges $T_{trailing}$ (keep $\alpha$ and $\beta$ unchanged) are plotted in figure~\ref{fig:force}(a), it is clearly shown that the thrust is a pulse, and the pulse terminates before 0.1~ms when the charge is finally dissipated. In \cite{Thomas2019}, the induced transient velocity measured by hot--wire was also found to be a pulse. Plotting the experimentally measured velocity curve (red dashed line) together with the calculated normalized thrust under voltage waveform extracted from~\cite{Thomas2019} in figure~\ref{fig:force}(b), we can find that the peak time of velocity and the thrust are in good agreement~\cite{Corkereview,Leonov2016}.
    
    (2) The theoretical upper limit of the thrust $F_{limit}$. The maximum peak thrust can be deduced from formula~(\ref{eqs.thrust_max}) assuming $t_{peak}=0$, which gives $F_{limit}=\varepsilon q_0V_m/2l$. Note that this value can't be achieved as the time required for the potential gap between the electrode and potential increase to 600~V is always non-zero. Nevertheless the physical upper limit of the thrust of the pulsed-DC actuator is estimated.
    
    \begin{figure}[t!h!]
    	\epsfxsize=\columnwidth
    	\begin{center}
    		\epsfbox{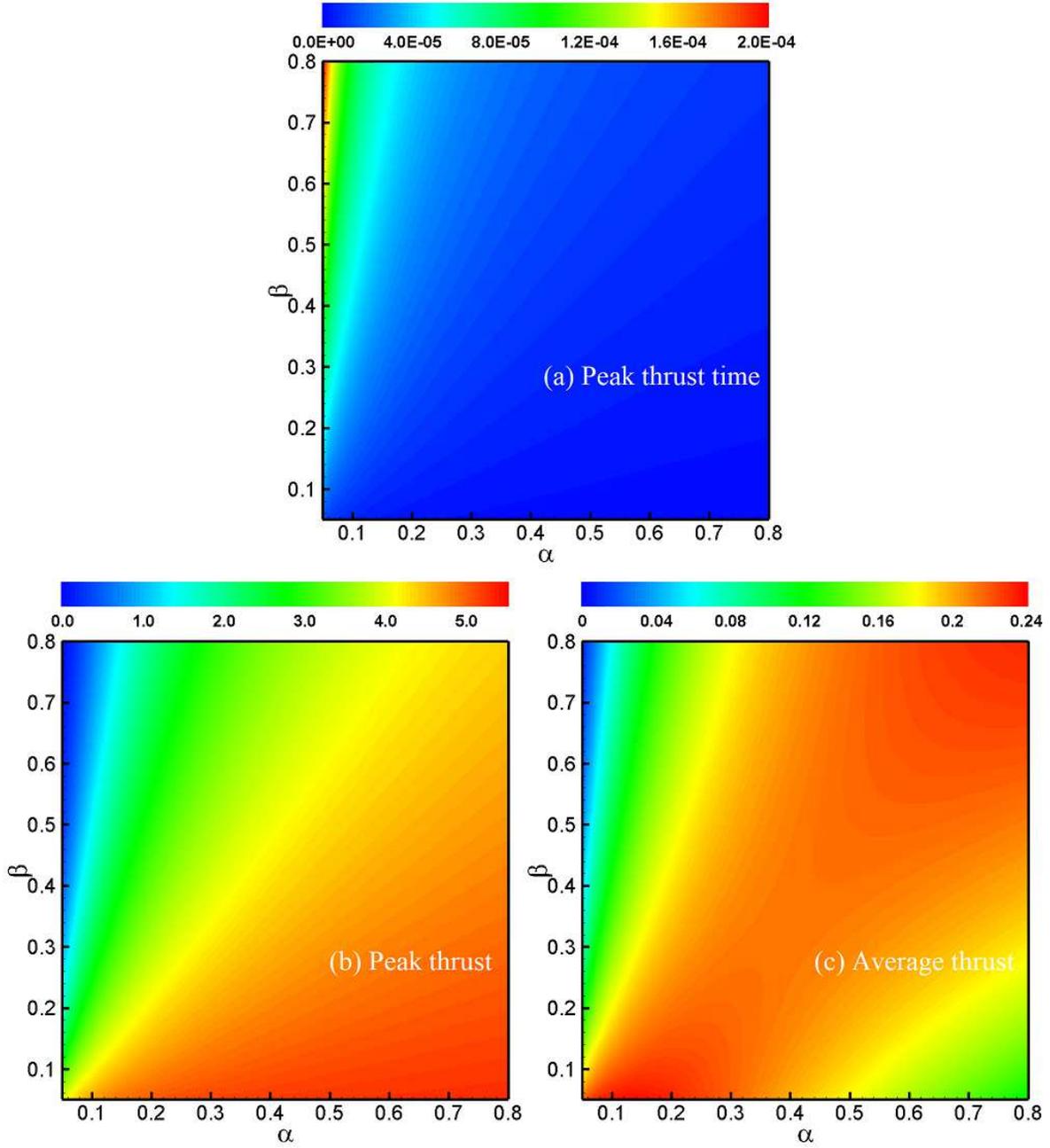}	
    		\caption{The function of peak thrust time $t_{peak}$, peak thrust $F_{peak}$ and average thrust $F_{mean}$ during trailing edge stage derived from equation~(\ref{eqs.bodyforce2}) as $\alpha$ and $\beta$. The period of applied voltage $T_v$ in equation~(\ref{eqs.bodyforce2}) equals to $T_{trailing}$ (0.6~ms).}
    		\label{fig:forceave}
    	\end{center}
    \end{figure}
    
    (3) The value and time moment of the peak thrust. These two parameters, $F_{peak}$ and $t_{peak}$, decide how and when the actuator can affect the air flow. Both parameters can be written as functions of trailing edge $T_{trailing}$. $F_{peak}$ appears when $V(t)=V_m-\Delta V_c=V_m-600$, with equation~(\ref{eqs.voltage}) we have following approximate expression for $t_{peak}$:
    
    \begin{equation}\label{eqs.falling_analy}
    t_{peak}\approx\frac{600}{V^{'}(0)}=\frac{600}{\alpha V_m\mathrm{ln}(\alpha V_m)/\beta+(1-\alpha)V_m\mathrm{ln}((1-\alpha)V_m)}T_{trailing}
    \end{equation}
    \noindent
    substituting the expression of $t_{peak}$ into formula~(\ref{eqs.thrust_max}) we can obtain the value of peak thrust $F_{peak}$:
    
    \begin{equation}\label{eqs.forcemax_analy}
    F_{peak}\approx\frac{\varepsilon q_0}{2l}V_m \mathrm{e}^{-\frac{600}{\alpha V_m\mathrm{ln}(\alpha V_m)/\beta+(1-\alpha)V_m\mathrm{ln}((1-\alpha)V_m)}\frac{T_{trailing}}{\Delta\tau_q}}
    \end{equation}

    In equations~(\ref{eqs.falling_analy}) and (\ref{eqs.forcemax_analy}), there are five key factors: charge density $q_0$, voltage amplitude $V_m$, time duration of voltage trailing edge $T_{trailing}$ and voltage profile parameters $\alpha$ and $\beta$. The charge density can be enlarged by shorten the rising slope or raise the voltage amplitude.     
    
    According to equation~(\ref{eqs.falling_analy}), the time of peak thrust $t_{peak}$ is a linear function of $T_{trailing}$, while from equation~(\ref{eqs.forcemax_analy}) we see the value of the peak thrust $F_{peak}$ decreases exponentially with $T_{trailing}$. Thus, shorter voltage trailing edge, in general, results in higher value and earlier appearance of the peak thrust. Here we consider two extreme cases as examples to illustrate the influence of $T_{trailing}$: (i) if the trailing edge is too long ($T_{trailing}>100~\mu s$), the transient thrust would be too small to affect the flow, which is often the case of positive period in an acSDBD; (ii) if the trailing edge is too short ($T_{trailing}<1~\mu s$), although higher transient thrust can be achieved, the time moment of peak thrust could be even smaller than the characteristic fluid response time (1~$\rm\mu s\sim1~\rm ms$), thus the effect of active flow control could still be weak, which is often the case of a nanosecond pulsed SDBD.
    
    The detailed profile of the trailing edge also affects the performance. $\alpha$ and $\beta$ are two parameters characterizing the decay rate of voltage. The contour maps of peak thrust time, peak thrust and average thrust as functions of $\alpha$ and $\beta$ are plotted in figure~\ref{fig:forceave} according to equation~(\ref{eqs.falling_analy}) and (\ref{eqs.forcemax_analy}), the value of $T_{trailing}$ is fixed as 0.6~ms. As can be seen, the larger $\alpha$ (or the smaller $\beta$), the faster decay rate, thereby the shorter time to decrease by 600~V (so-called peak thrust time, see figure~\ref{fig:forceave}~(a)), while peak thrust value shows the opposite trend (figure~\ref{fig:forceave}~(b)). The maximum average thrust appears when both $\alpha$ and $\beta$ are small or large (figure~\ref{fig:forceave}~(c)), which means that if the amplitude of one sub-voltage (the first term in right side of equation~(\ref{eqs.voltage})) is large, its trailing edge ($\beta T_{trailing}$) should also be chosen larger to maintain a longer duration for high thrust, and vice versa.
	    
    \section{Conclusions}
    The pulsed--DC SDBD (or the tailored Fast-Rise-Slow-Decay SDBD, FRSDBD) actuators are studied with the help of a 2D fluid model and an analytical model. The discharge characteristics during the voltage rising (pulse) stage, and thrust generation during the voltage decaying stage, as well as the factors influencing the thrust are analyzed in detail. Following conclusions can be drawn:
    
    (1) The fast voltage rising stage is responsible for plasma channel formation and charge production. The thrust in this stage is negative. The breakdown takes place when the applied voltage is 1040~V, a plasma channel with the length of 4.3~mm and thickness of 50~$\mu$m is formed. A current spike appears at breakdown moment then drop within 490~ns, much shorter than the DC--pulse width. The charge density in the channel is negative, while the $x$--component of electric field $E_{xl}$ is positive. Shorter rising slopes lead to higher electron density (thus charge density). No positive thrust is generated in this stage.
    
    (2) During the slow voltage decaying stage, the $x$--component of electric field $E_{xl}$ changes its direction and positive thrust starts due to collisions of negative ions and neutral molecules. The thrust reaches its maximum when the applied voltage decreases by about 600~V from the peak value; then the evolution of thrust in subsequent times is dominated by charge due to the fast deposition of negative ions on dielectric surface. 
    
    (3) The lasting time of the trailing edge $T_{trailing}$, the decay rate of applied voltage (described by two parameters $\alpha$ and $\beta$), and voltage amplitude $V_m$ are the main factors deciding the peak thrust time and value. Analytical expressions are formulated for the peak thrust value and time moment based on these parameters and input from 2D simulations. 
    
    (4) The time moment of peak thrust $t_{peak}$ increases linearly with the length of the trailing edge $T_{trailing}$, while peak thrust value $F_{peak}$ is an exponentially--decreasing function of $T_{trailing}$. 
    The upper limit of $F_{peak}$ value is obtained. 
    A 2D contour map describing the peak thrust value and time, and the average thrust is drawn. Our results show that, smaller voltage decay rate at the beginning results in a delayed peak thrust time and a smaller peak thrust value. The average thrust in one duty cycle decreases with the decay rate of applied voltage.
    
    The conclusions drawn above allow us to flexibly design the FRSD waveformes for an SDBD actuator according to the requirements of active flow control in different application conditions.

	\section*{Acknowledgements}
	The work was partially supported by the National Natural Science Foundation of China (No. 51790511, 51907204, 91941105, 91941301) and the National Numerical Windtunnel Project NNW2018-ZT3B08. The authors are thankful to the young research group in Atelier des Plasmas for fruitful discussions.
		
	\section*{References}
	\bibliography{Libaray_ZHU}
	\bibliographystyle{ieeetr}

\end{document}